\documentclass[12pt,english,letterpaper]{article}

\usepackage{babel}
\usepackage{graphicx}
\usepackage{amsfonts}
\usepackage{latexsym}
\usepackage[round]{natbib}

\begin{document}
\pagestyle{myheadings}
\renewcommand{\thefigure}{\arabic{figure}}

\begin{center}

\bf
\Large

Gravitational Lensing in the metric theory proposed by Sobouti

\large

Tula Bernal $\footnote{tbernal@astroscu.unam.mx}$ \& Sergio Mendoza 
$\footnote{sergio@mendozza.org}$

\normalsize

Instituto de Astronom\'{\i}a, Universidad Nacional Aut\'onoma de M\'exico,
Ciudad Universitaria, M\'exico.

\end{center}

\normalsize

\begin{abstract}

 Recently, Y. Sobouti (2007) has provided a metric theory $f(R)$ that
can account for certain dynamical anomalies observed in spiral galaxies.
Mendoza \& Rosas-Guevara (2007) have shown that in this theory there is an
extra-bending as compared to standard general relativity. In the present work
we have developed in more specific detail this additional lensing effect
and we have made evaluations of the $\alpha$ parameter used in the model
adjusting the theory to observations in X-rays of 13 clusters of galaxies
with gravitational lensing (\cite{hoekstra07}).

\end{abstract}

\section{Introduction}

Some modified theories of gravity are an actual alternative to
the dark matter paradigm, which in spite of its success to
explain astrophysical problems like the flat rotation curves
of spiral galaxies, the gravitational lensing by clusters of
galaxies and the formation of structure in the early Universe,
is not the ultimate word of the story.

In the different alternatives we find the modifications to the Hilbert-Einstein
action, which introduce an arbitrary function of the Ricci scalar
\(R\) in the action. This modification is relevant, because there's no
exist an \it a priori \rm reason to restrict the gravitational action to
\(R\) and has the advantage to be a metric theory of gravitation that
carry out with several physical principles and at the same time accounts
to the energy-momentum conservation, as well as the equations of the
theory are invariants.

In this sense, Sobouti (2007) has developed a metric theory $f(R)$ that
reproduces flat rotation curves in different spiral galaxies (with
$f(R)=R^{1-\alpha /2}$ and $\alpha \ll 1$), reproduces the Tully-Fisher
relation, converges to a version of MOND in the weak field approximation
and the resulting differential field equations are of the second order.

\section{Modified field equations}

The model introduces the following modification to the
Hilbert-Einstein action (\cite{capozz02}; \cite{capozz03}):

\begin{equation}
  S_g = - \int{ \frac{1}{2} f(R) \sqrt{-g } \rm{d}\Omega } \label{ionflux}.
\end{equation}

Variation of the action $S=S_g+S_m$ with respect to the metric
$g_{\mu \nu}$ gives the following field equations (\cite{capozz03}):

\begin{equation}
  f'(R) R_{\mu \nu} - \frac{1}{2} g_{\mu \nu} = T_{\mu \nu} +
  (f'(R))_{;\mu \nu} - {(f'(R)) _ {;\lambda}} ^{\lambda} g_{\mu \nu} \label{ionflux}.
\end{equation}

For an application of this theory to different galactic systems,
the metric is chosen like a Schwarzschild-like one given by (\cite{cognola05};
 \cite{sobouti07}):

\begin{equation}
  ds^2 = B(r)dt^2 - A(r)dr^2 - r^2 \left( d\theta^2 + \rm{sin}^2\theta
  d\varphi^2 \right) \label{ionflux}.
\end{equation}

If the function $f$ differs not too much from general relativity at
first order of approximation, the exterior solution ($\rho=0=p$) to the
field equations is obtained with the assumption that $v = df/dR$ is a
parametric function $v(r,\alpha)$ with $\alpha$ a real number such that
$\alpha\ll 1$ (for $\alpha = 0$ the solution converges to G.R.). Proposing a solution of the form $v \propto r^\beta 
\rm{ , } \beta \in \rm{R}$, results in $v \simeq r^\alpha$ and the
solutions for the metric are (\cite{sobouti07}):

\begin{equation}
  \frac {1} {A(r)} = \frac {1}{1-\alpha} \left[ 1 - \left( \frac {s}{r}
                     \right) ^ {1 - \alpha / 2} \right]; \rm{ }\rm{ }
  B(r) = \left( \frac {r}{s} \right) ^ \alpha \frac {1}{A(r)} \label{ionflux};
\end{equation}

\noindent where $s:=2GM$ is the Schwarzschild radius. For $\alpha\ll 1$, $f(R)$ is
given by:

\begin{equation}
  f(R) = \left( 3\alpha \right) ^ {\alpha / 2} s ^ {-\alpha} R ^ {1 - 
         \alpha / 2} \simeq R \left[ 1 - \frac {\alpha}{2} \rm{Ln} \left(
         s^2 R \right) + \frac {\alpha}{2} \rm{Ln}(3 \alpha) \right] \label{ionflux}.
\end{equation}

\section{The deflection of light and gravitational lensing}

Mendoza \& Rosas-Guevara(2007) have shown that the deflection angle
$\beta$ for a spherically symmetric mass distribution in this metric
theory (cf. {eq.3}), is given by:

\begin{equation}
  \beta = \pi \left[ \frac {2 \sqrt {1-\alpha}} {2-\alpha} - 1 \right]
  + 2 \sqrt{1-\alpha} \left( \frac {s}{r_m} \right) ^ {1-\alpha /2} \label{ionflux},
\end{equation}

\noindent where $r_m$ is the impact parameter (the closest approach to the lens). When
$\alpha = 0$, the deflection angle equals the Einstein's one: $\beta _ E = 
2s / r_m$.

For an application in gravitational lensing for a symmetric
object (e.g. a cluster of galaxies) with mass $M$, the lens equation is given
by (\cite{mendoza07}):

\begin{equation}
  \Theta^2 - \Theta (\Phi + C_1) - C_2 \Theta ^ {\alpha / 2} = 0 \label{ionflux} ,
\end{equation}

\noindent where: $C_1 := \frac{\pi}{\theta_E} \frac{D_{LS}}{D_S} \left[ \frac
               {2 \sqrt{1-\alpha}}{2-\alpha} - 1 \right]$;
$C_2 := \theta _E ^{-\alpha/2} {\left( \frac{1}{2} \frac{D_S}{D_{LS}}
        \right)}^{-\alpha/2} \sqrt{1-\alpha}$;  
       $ \Theta_E = \sqrt {\frac{4GM}{D_S} \frac{D_{LS}}{D_L}} $;  
       $\Theta := \theta / \theta_E$;  
       $\Phi := \gamma / \theta_E$; and $\gamma$ is the angular position of the source,
       $\theta$ is the angular position of the image, $D_S$, $D_L$ and $D_{LS}$ are the
       observer-source, observer-lens and lens-source distances, respectively.

The solution to the lens equation at first order, with
$\Theta_1$ a small linear perturbation to the solution of general
relativity, $\Theta_0 = \frac{1}{2} \left( \Phi \pm \sqrt{\Phi^2 + 4}
\right)$, is given by (\cite{mendoza07}):

\begin{equation}
  \Theta_1 = \frac {C_1 \Theta_0 + C_2 \Theta_0 ^{\alpha/2} - 1}
  {2 \Theta_0 - \Phi - C_1 - \frac{\alpha}{2} C_2 \Theta_0 ^ {\alpha/2-1}} \label{ionflux}.
\end{equation}

With this perturbation the magnification factor is (\cite{mendoza07}):

\begin{equation}
  \mu = \rm{det} \left[ J(\theta) \right] ^ {-1} = \left| \frac{\Theta}
  {\Phi} \rm{ } \frac{d\Theta}{d\Phi} \right| \simeq \frac{\Theta_0 +
  \Theta_1} {\Phi} \rm{ } \frac{d\Theta_0}{d\Phi} \label{ionflux},
\end{equation}

\noindent where $J(\theta)$ is the jacobian of the transformation $\theta \rightarrow
\gamma$ (from the lens plane to the source plane).

For $\rm{det} \left[ J(\theta) \right] = 0$ (when the magnification factor $\mu$
diverges), the critical curve for $\Phi=0$ and for $\theta_0^+(0) = \theta_E$, is given by:

\begin{equation}
  \theta_C^+ = \frac {1 + C_2 \left( 1 - \frac{\alpha}{2} \right)} {2 - C_1
  - \frac{\alpha}{2} C_2 } \theta_E \label{ionflux},
\end{equation}

\noindent and for $\theta_0^-(0) = 2\pi - \theta_E$ by:

\begin{equation}
  \theta_C^- = \frac { 2 {\left( \Theta_0^-\right)}^2 - 1 + C_2 {\left(
  \Theta_0^-\right)}^{\alpha/2} \left( 1 - \frac{\alpha}{2} \right)}
  {2\Theta_0^- - C_1 - \frac{\alpha}{2} C_2 {\left( \Theta_0^-\right)
  }^{\alpha/2 - 1} } \theta_E \label{ionflux}.
\end{equation}

From the symmetry of the problem we have $(2\pi - \theta_C^-) = \theta_C^+ :=
\theta_C$, and we can take $\theta_C$ as the equivalent to the
Einstein angle $\theta_E$, i.e. an "Einstein ring" in this theory
have an aperture of $\theta_C$, and $\theta_C = \theta_E$ for $\alpha = 0$.

\section{Approximation method to $\alpha$}
\label{approximation}

Since these results are derived from a metric with spherical symmetry,
we use observations in X-rays of gravitational lensing by 13 clusters of
galaxies that approximate well to this symmetry (see \cite{hoekstra07}
and references therein). In order to estimate the parameter $\alpha$, we
assume that the estimated values of the Einstein angle from the observations
approximates $\theta_C$. We use this as a first way to mesure the order of
magnitude to $\alpha$.

The cosmological model used is: $H_0 = 70 \rm{ km} \rm
{ s}^{-1} \rm{ Mpc}^{-1}$, $\Omega_m = 0.3$ and $\Omega_\Lambda = 0.7$.
To calculate the Einstein angle we use the virial mass of the system with
the mean velocity dispersion observed in a particular system (\cite{borgani99};
\cite{girardi01}).

Using this method we can solve numerically the equation (10) to obtain $\alpha$.

\section{Results}

   Sobouti's \( f(R) \) theory is a first approximation to a modified
metric theory of gravity in order to account for phenomena usually
adscribed to dark matter.  However, there is a caveat with Sobouti's
approach.  The mass dependency of \( \alpha \) destroys one of the most
important facts of an ordinary action: it should be invariant under the
changes of sources.   Also, this mass dependency means that the function \(
f(R) \) varies from source to source. However, as mentioned by 
Sobouti~(2007), perhaps we need to re-think whether this postulate 
has to be accepted when modifications to standard general relativity 
are done.

  The way Sobouti (2007) calculated the value for the parameter \( \alpha
\) was using a spherically symmetric metric, applied to spiral galaxies.
This is of course wrong since these galaxies are far from spherical
symmetry.  This is the reason as to why we decided to callibrate the \(
\alpha \) parameter using spherical symmetric astrophysical systems:
cluster of galaxies.  From the many cluster of galaxies known we have
chosen a set of very well spherically symmetric objects.

  To summarize, our main results are as follows:

\begin{itemize}

\item  The caustic in the source plane is the point colinear
with the lens and the observer. This is an expected result since the metric
is a Schwarzschild-like one for a spherically symmetric matter distribution.

\item The $\alpha$-mass relation from Sobouti (2007) is given by $\alpha = \alpha_0 {\left( \frac{M}{M_\odot} \right)}^{1/2}$.

\noindent Our best linear fit gives a value of $\alpha_0 = (3.5 \pm 0.3)$x$10^{-9}$
(see Fig. 1), that differs in three orders of magnitude from the value obtained
by Sobouti (2007), $\alpha_0 = (3.07\pm0.18)$x$10^{-12}$, for 31 spiral
galaxies with $\alpha = 2 v_\infty ^ 2 / c^2 $ ($v_\infty$ is the
asymptotic constant tangential speed of the galaxy).

\begin{figure}
\includegraphics[width=12cm]{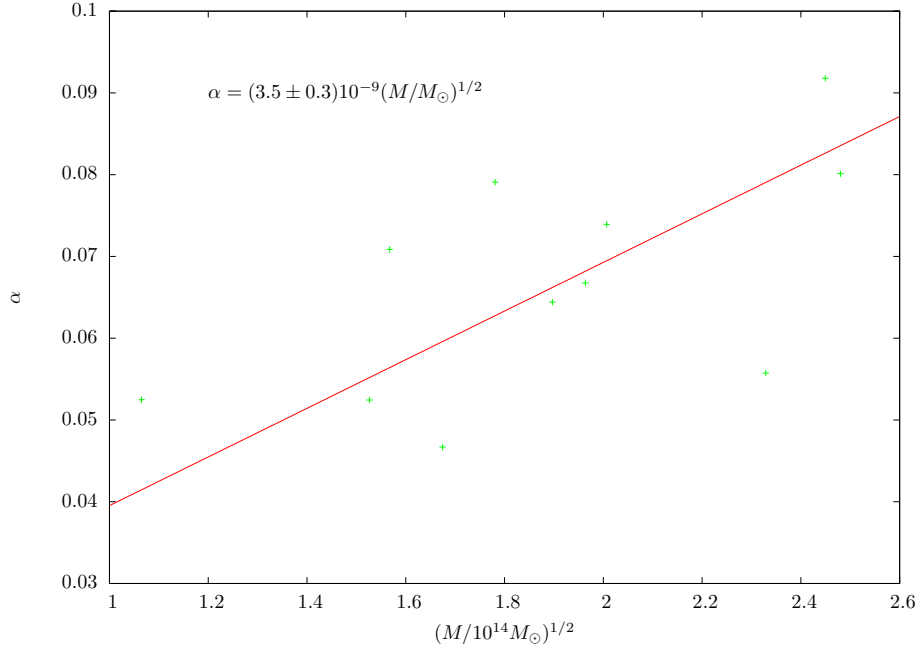}
\centering
\caption{Plot of $\alpha$ vs. $(M/10^{14}M_\odot)
       ^{1/2}$. The best linear fit is shown by the line.}
\end{figure}

\item The log-log plot of $\alpha$ vs. $M/M_\odot$, leads the power law
fit: $ \alpha = \left[ (1.47 \pm 0.51) \rm{ x } 10^{-5} \right]
\left( \frac{M} {M_\odot} \right) ^ {0.25} $, i.e. a relation
of the form: 

\begin{equation}
  \alpha (M) = \alpha_1 \left( { \frac{M} {M_\odot} } \right)^{1/4} \label{ionflux}.
\end{equation}

\begin{figure}
\includegraphics[width=12cm]{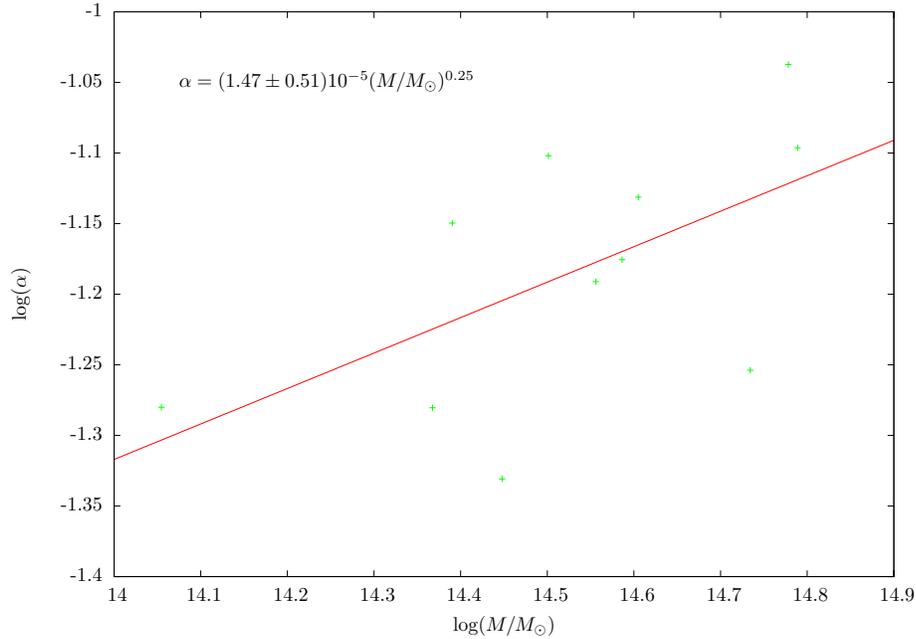}
\centering
\caption{Plot of $\rm{log}(\alpha)$ vs.
      $\rm{log}(M/M_\odot)$. The best linear fit is shown by the line.}
\end{figure}

\item Although the calculations are only a first approximation
to the order of magnitude to the $\alpha$ parameter, this is the correct way for
evaluating it, since the distribution of barionic matter approximates to a
spherical symmetric one in clusters of galaxies.

\item One could be tempted to use the results also for quite ``spherical''
elliptical galaxies.  However, this approximation fails because 
the Einstein angle in these galaxies is very small and so the 
variables $C_1$ and $C_2$ (cf. {eq.7}) diverge.

\item \it Future work \rm:  In section \ref{approximation} we approximated
the value \( \alpha \) by using the estimated Einstein angles and equate
them to \( \theta_C \).  This is a first approximation to
the real value of \( \alpha \) and the only way to correct this is to 
develope a complete theory of
gravitational lensing for the metric theory proposed by Sobouti. In
this way, we intend to determine with much precision the exact value
for the $\alpha$ parameter, and check whether this theory can account
for different astrophysical problems.

\end{itemize}

\section{Acknowledgments}
T. Bernal acknowledges financial support from CONACyT (207529) and
S. Mendoza acknowledges financial support from DGAPA-UNAM (IN119203)
and (IN11307-3).

\end{document}